\documentclass[prd,aps,floatfix,notitlepage,reprint,superscriptaddress,twocolumn,nofootinbib]{revtex4-1}
\pdfoutput=1
\usepackage{graphicx,bm,graphics,amsmath,bm,natbib,multirow,hyperref,float,fancyhdr,amssymb,amsfonts}
\usepackage[usenames,dvipsnames]{color}
\usepackage[caption = false]{subfig}
\hypersetup{colorlinks=true,citecolor=blue,linkcolor=blue}

\newcommand{\be}{\begin{equation}}
\newcommand{\ee}{\end{equation}}
\newcommand{\zeff}{z_\mathrm{eff}}

\begin{document}

\title{Anisotropic effective redshift and evolving clustering amplitude}
\author{Andrej Obuljen}
\affiliation{Waterloo Centre for Astrophysics, University of Waterloo, 200 University Ave W, Waterloo, ON N2L 3G1, Canada}
\affiliation{Department of Physics and Astronomy, University of Waterloo, 200 University Ave W, Waterloo, ON N2L 3G1, Canada}
\author{Will J. Percival}
\affiliation{Waterloo Centre for Astrophysics, University of Waterloo, 200 University Ave W, Waterloo, ON N2L 3G1, Canada}
\affiliation{Department of Physics and Astronomy, University of Waterloo, 200 University Ave W, Waterloo, ON N2L 3G1, Canada}
\affiliation{Perimeter Institute for Theoretical Physics, 31 Caroline St. North, Waterloo, ON N2L 2Y5, Canada}

\begin{abstract}
A typical galaxy survey geometry results in galaxy pairs of different separation and angle to the line-of-sight having different distributions in redshift and consequently a different effective redshift. However, clustering measurements are analysed assuming that the clustering is representative of that at a single effective redshift. We investigate the impact of variations in the galaxy-pair effective redshift on the large-scale clustering measured in galaxy surveys. We find that galaxy surveys spanning a large redshift range have different effective redshifts as a function of both pair separation and angle. Furthermore, when considering tracers whose clustering amplitude evolves strongly with redshift, this combination can result in an additional scale-dependent clustering anisotropy. We demonstrate the size of this effect on the eBOSS DR16 Quasar sample and show that, while the impact on monopole is negligible, neglecting this effect can result in a large-scale tilt of $\sim 4\%$ and $\sim40\%$ in quadrupole and hexadecapole, respectively. We discuss strategies to mitigate this effect when making measurements.
\end{abstract}

\maketitle

\section{Introduction}\label{Intro}
Large-scale galaxy clustering is one of the most important and promising observations from which we can extract cosmological information about the late time Universe. The success of recently finished Sloan Digital Sky Survey IV (SDSS-IV; \cite{SDSSIV_Blanton}) extended Baryon Oscillation Spectroscopic Survey (eBOSS; \cite{eBOSS_Dawson}), together with BOSS \cite{BOSS_Dawson}, has directly demonstrated their constraining power \cite{Alam_BOSS,Alam_eBOSS}. In the near future we expect measurements at an even higher level of accuracy with surveys including DESI\footnote{\href{https://www.desi.lbl.gov}{https://www.desi.lbl.gov}}\cite{DESI} and Euclid\footnote{\href{http://sci.esa.int/euclid/}{http://sci.esa.int/euclid/}}\cite{Euclid}, probing order-of-magnitudes larger volumes. One of the main goals of these upcoming surveys is to measure the two-point galaxy clustering at an unprecedented precision. 

The observed large-scale galaxy clustering amplitude is approximately linearly biased with respect to the clustering of underlying matter \cite{Kaiser_bias} (see \cite{Wechsler_review,Bias_review} for reviews). Additionally, the observed galaxy clustering is anisotropic due to the effect of Redshift-Space Distortions (RSD) \cite{Kaiser,Hamilton92,Hamilton98}, and the Alcock-Pacynski effect \cite{AP}. RSD allow us to constrain the large-scale structure growth rate, test General Relativity (GR) and constrain sum of the neutrino masses \cite{Guzzo,PercivalWhite,Lesgourgues}. The large-scale clustering in redshift-space is in principle straightforward to model within linear theory under the plane-parallel approximation on sub-horizon scales. Within this approximation one assumes parallel line-of-sight (LOS) directions to each object in a pair. However, realistic surveys come with additional large-scale effects that can limit the simple interpretation of clustering measurements. These effects can be roughly separated into wide-angle, window and relativistic effects. 

Wide-angle effects become significant whenever the plane-parallel approximation is violated, i.e.\ when objects pair separations become comparable to the their distances from the observer \cite{Raccanelli+2010, Samushia+2011,YooSeljak,Castorina_wide}. This happens, for instance, when considering shallow survey and large-scales.
As wide-angle effects are related to individual pairs, they can be separated from the distribution of the pairs in a survey as a whole.

A typical galaxy redshift survey only covers a fraction of the sky with varying redshift selection function, which results in a survey window function. The survey window function quantifies which pair separation vectors are available. In Fourier space, the power spectrum prediction is convolved with the window function in order to account for the survey geometry \cite{Window_Beutler,Wilson}. The correlation function of a galaxy survey is multiplied by the window such that the expected number of pairs as a function of their orientation with respect to the LOS changes with the amplitude of the pair separation. At larger separations this distribution can be significantly different from an isotropic distribution that is evenly distributed in $\mu$, where $\mu$ is the cosine of the angle a separation vectors makes with the LOS \cite{Samushia+2011, YooSeljak}. This is usually included when calculating the correlation function by integrating over $\mu$ after normalising by the expected number of pairs in each $\mu$-bin. 

Beyond the standard Kaiser approximation which assumes Newtonian dynamics a fully relativistic description of the observed quantities becomes necessary \cite{Yoo2009,Yoo2010,BonvinDurrer,ChallinorLewis,Lepori}. On scales comparable to the horizon, the relativistic effects become substantial and the observed clustering deviates strongly from a simple theoretical descriptions. However, the window function convolution is still needed in addition to including relativistic effects in the theoretical modeling.

One of the motivations for utilizing the large-scales is to probe the effects of primordial non-gaussianities (PNG) which leave a scale-dependent signature on the large-scale galaxy bias \cite{Dalal_fnl,MatarreseVerde,Slosar_fnl}. Currently the best constraints made using this signal come from the eBOSS quasar sample \cite{Castorina_fnl} which cover a wide redshift range allowing us to probe the largest scales. Another motivation for probing the large scales is to test GR on (sub)horizon scales using galaxy surveys \cite{Yoo2012} and disentangle the GR effects from the PNG effects \cite{Wang_Beutler}. Surveys such as DESI are forecast to detect relativistic effects \cite{Beutler_DiDio}. Thus it is crucial to understand and account for all possible systematic effects in order not to bias our interpretation of clustering measurements. 

In general, galaxy surveys cover a wide redshift range across which the properties of galaxy populations will change. These include the amplitude of the RSD and matter clustering strength, which evolve due to the evolution of the Universe, and galaxy bias which evolves as the average properties of the galaxies observed change. Such effects can, in principle, be easily included in models of wide-angle and relativistic effects, resulting in extra redshift-dependent terms in the relevant integrals. However, they couple with the window function leading to a pernicious complication, which is most easily understood by thinking about the effective redshift of a sample. Standard clustering analyses usually assume a single effective redshift at which the model is compared to the measurements. However, for a sample whose window function changes the distribution of pairs in $\mu$ with redshift, the effective redshift will vary with $\mu$. If furthermore, we have a sample whose galaxy clustering strength and the strength of the RSD also vary with redshift, then we will pick up a tilt in the measured clustering moments with respect to that of a sample whose clustering properties do not change across the sample. 

In this paper we consider the impact of the survey geometry in the presence of clustering evolution on the interpretation the anisotropic galaxy clustering. As we are only interested in this effect, we make some simplifying assumptions: We do not include relativistic or wide-angle effects. Furthermore, as we are mostly interested in the large-scales clustering, we work at the level of linear RSD theory. To demonstrate the size of the effect using a real sample, we consider the public eBOSS quasar (QSOs) sample as it spans the largest redshift range ($0.8<z<2.2$) making it ideal to probe the largest scales \cite{Ross-eboss-cats}. Additionally, the QSO sample has the strongest clustering evolution compared to other eBOSS tracers.

This paper is organized as follows. In \S\ref{theory} we review the linear RSD theory of the large-scale galaxy clustering and its main observables defined at a single redshift. In \S\ref{Effectivez} we define the anisotropic effective redshift and discuss its effect on the clustering predictions. In \S\ref{data} we describe the eBOSS data and measure the anisotropic $\zeff$ using the random catalogs of the QSO sample. In \S\ref{results} we show the impact of neglecting the anisotropic $\zeff$ on the clustering multipoles both in configuration and Fourier space. Finally, we discuss our results, consider the impact on cosmological parameters and present our conclusions in \S\ref{discussion}.

In this paper we assume the cosmological parameters from Planck 2015 results \cite{planck15}, as implemented in \texttt{nbodykit} package \cite{nbodykit}.

\section{Linear RSD clustering}\label{theory}
In linear theory on subhorizon scales the observed redshift-space galaxy power spectrum at redshift $z$ can be described by the  Kaiser RSD model \cite{Kaiser}:
\be
\label{eq:Pkmu}
P(k,\mu,z) = D^2(z)\left(b(z)+f(z)\mu_k^2\right)^2 P_m(k,0),
\ee
where $D$ is the linear growth factor, $b$ is the linear galaxy bias, $f=d \ln D(a)/d \ln a$ is the logarithmic growth rate, $\mu_k=k_\parallel/k$ and $P_m(k,0)$ is the linear matter power spectrum at $z=0$. This form of anisotropic power spectrum can be expressed in terms of the Legendre multipoles as:
\be
P_{\ell}(k)\equiv\frac{2\ell+1}{2}\int_{-1}^1 P(k,\mu_k)\mathcal{L}_\ell(\mu_k)d\mu_k,
\ee
where $\mathcal{L}_\ell$ is the Legendre polynomial of order $\ell$. In linear RSD theory the only non-zero multipoles are $\ell=0,2,4$ which are given by:
\be
\begin{split}
P_{0}(k) &=\left(b^2+\frac{2}{3} f b + \frac{1}{5}f^2\right)P_m(k),\\
P_{2}(k) &=\left(\frac{4}{3} b f+\frac{4}{7}f^2\right)P_m(k),\\
P_{4}(k) &=\frac{8}{35}f^2 P_m(k).
\end{split}
\ee

The redshift-space galaxy clustering can be defined by its Fourier transform in configuration space, the two-point correlation function $\xi(r,\mu)$, which depends on the galaxy pair separation $r$ and the cosine of the angle that the separation vector makes with the LOS  $\mu$. As in the case of the power spectrum, the correlation function $\xi(r,\mu)$ can be expressed in terms the Legendre multipoles by:
\be
\xi(r,\mu)=\sum_\ell \xi_\ell(r)\mathcal{L}_\ell(\mu),
\ee
where the correlation function multipoles $\xi_\ell$ are related to the power spectrum multipoles through:
\be
\label{eq:xiell}
\begin{split}
\xi_\ell (r) &= \frac{i^\ell}{2\pi^2} \int k^2 dk P_\ell (k) j_\ell (kr),\\ 
P_\ell (k) &= (-i)^\ell (4\pi) \int r^2 dr \xi_\ell (r) j_\ell (kr), 
\end{split}
\ee
where $j_\ell$ is the spherical Bessel function.

\section{Anisotropic effective redshift}\label{Effectivez}
Galaxy samples targeted by galaxy redshift surveys typically cover a wide redshift range. In order to measure clustering on large scales, they are often analysed in wide redshift bins, and the clustering is interpreted as being representative of that at a single effective redshift. It is at this redshift that the model is calculated in order to compare to the measurements. Some quantities and hence the measurements made from them, are invariant with redshift such as the shape of the linear power spectrum, and are not sensitive to the exact value of the effective redshift chosen. In contrast, the RSD signal varies with redshift and the effective redshift calculation is important for these measurement. The survey's effective redshift based on galaxy pairs is defined as \cite{Hou,Neveux}:
\be
\label{eq:zeff}
z_\mathrm{eff}\equiv\frac{\sum_{i,j}w_i w_j(z_i+z_j)/2}{\sum_{i,j}w_i w_j},
\ee
where $w_i$ are the weights applied to data in order to account for the observing and targeting systematics. The effective redshift is typically measured by summing over all pairs with separations in the range $s=[20,160]\, h^{-1}\mathrm{Mpc}$, which gives the value that is used as the effective redshift of the full sample in the clustering analysis. The $z_\mathrm{eff}$ associated with a clustering measurement is commonly assumed to be the same irrespective of pair separation amplitude or orientation. 

Allowing for anisotropy in $\zeff$, the theoretical expectation of the measured clustering defined in Eq.\ \eqref{eq:Pkmu} at an effective redshift can also be described as:
\be
\label{eq:Pkmuzeff}
P(k,\mu,\zeff) = D^2(\zeff)\left(b(\zeff)+f(\zeff)\mu_k^2\right)^2 P_m(k,0),
\ee
where the effective redshift $\zeff(r,\mu)$ is both the function of pair separation $r$ and pair orientation $\mu$. This makes all the time varying quantities inherit the scale and orientation dependence in the model such that we have the following: $b(r,\mu)$, $D(r,\mu)$ and $f(r,\mu)$. In the following we will use the same definition as Eq.\ \eqref{eq:zeff}, but instead of only the pair separation, we will also consider the dependence of $\zeff$ on both $r$ and $\mu$.

\section{Application to eBOSS QSO sample}\label{data}
In this section we describe the QSO sample properties and the catalogs used in our analysis, compute the sample's anisotropic $\zeff$ and describe the clustering strength evolution of the QSO sample.

\subsection{eBOSS DR16 quasar sample}
The extended Baryon Oscillations Spectroscopic Survey\footnote{\href{https://www.sdss.org/surveys/eboss/}{https://www.sdss.org/surveys/eboss/}} (eBOSS) \cite{eBOSS_Dawson}, part of SDSS-IV \cite{SDSSIV_Blanton}, measured nearly $\sim900,000$ spectroscopic redshifts in the redshift range $0.6<z<2.2$ using three main galaxy samples: Luminous Red Galaxies (LRG), Emission Line Galaxies (ELG) and QSO. In the latest eBOSS Data Release 16 (DR16) the clustering measurements from these galaxy samples have been used to put constraints on cosmological parameters through the Baryon Acoustic Oscillation (BAO) and RSD measurements \cite{Bautista_BAO_RSD_LRG_eBOSS,Hector_BAO_RSD_LRG_eBOSS,Tamone_RSD_ELG_eBOSS,Arnaud_BAO_RSD_ELG_eBOSS,Hou,Neveux}. eBOSS measured spectroscopic redshifts to over $\sim300,000$ quasars in the redshift range $0.8<z<2.2$ and in the following we will use the latest, publicly available DR16 release of the QSO sample \cite{LSSCatalog}. The QSO sample is further split into the North and South Galactic Cap (NGC and SGC). We focus our analysis on the larger, NGC, part of the sample which contains $218,209$ quasars and covers a sky area of $2890\,\mathrm{deg^2}$.

There are two main reasons why we choose the eBOSS QSO sample to investigate the effect of the anisotropic $\zeff$ in this paper. One reason is that QSO sample covers the widest redshift-range among all the other tracers in SDSS. This makes this sample well suited to probe the largest scales, and susceptible to evolution across the sample. The other reason is that between the other (e)BOSS galaxy samples, QSO sample has the strongest redshift-evolution of the linear bias. In particular, the measured bias evolution $b_Q(z)$ evolves more strongly than the inverse growth factor $D(z)^{-1}$ which results in a stronger overall clustering strength evolution \cite{bQSO_Laurent,bQSO_Hector}.

To investigate the anisotropic $\zeff$ we make use of the random catalogs created for the eBOSS DR16 Quasar samples \cite{LSSCatalog}. These random catalogs were created to match the sampling and observational characteristics of the data catalog but without any clustering signal. Apart from the angular positions and redshifts, these catalogs also contain a number of weights for each object which account for observing and targeting systematics present in the data catalogs. We account for these weights when computing $\zeff$ by using the following total weights \cite{LSSCatalog}:
\be
\label{eq:wtot}
w_\mathrm{tot} = w_\mathrm{sys}\times w_\mathrm{cp}\times w_\mathrm{noz}\times w_\mathrm{FKP},
\ee
where $w_\mathrm{sys}$ are the imaging systematics weights, $w_\mathrm{cp}$ are the close pair weights, $w_\mathrm{noz}$ are the redshift failure weights and $w_\mathrm{FKP}$ are the standard FKP weights \cite{FKP}.
\begin{figure}[!ht]
\includegraphics[width=0.48\textwidth]{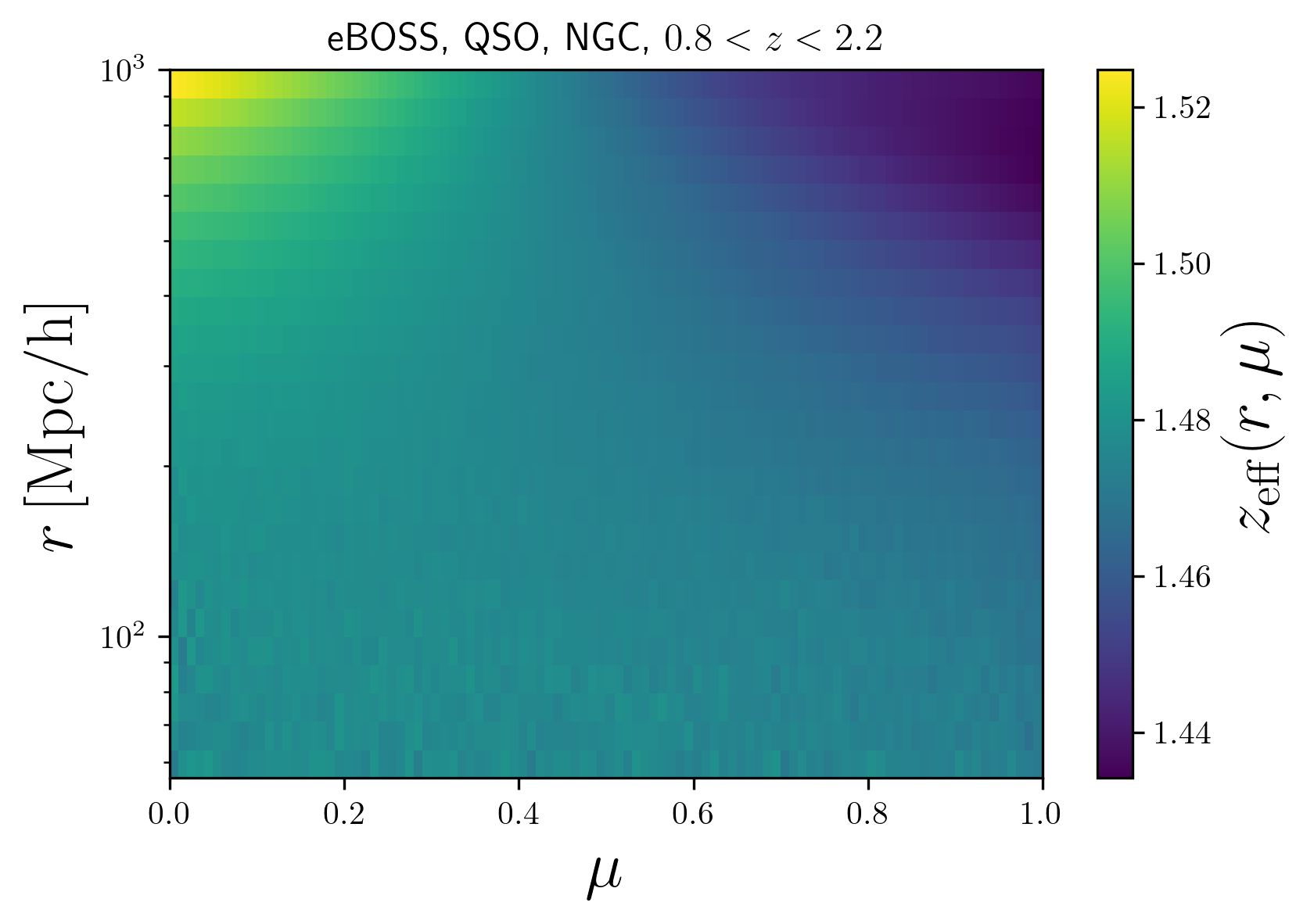}
\caption{Effective redshift as a function of 2D pair separation $(r,\mu)$ computed using the eBOSS QSO NGC random catalog. While isotropic on small pair separations, the effective redshift becomes anisotropic at separations $r\gtrsim300\, h^{-1}\mathrm{Mpc}$ due to the survey geometry.}
\label{fig:zeff}
\end{figure}
\subsection{eBOSS QSO anisotropic effective redshift}
We use the random catalogs described in the previous subsection to measure $\zeff(r,\mu)$, using Eq.\ \eqref{eq:zeff} and the weights defined in Eq.\ \eqref{eq:wtot}. For the pair separation $r$ we use $41$ logarithmic bins in the range $r=[10,1000]\, h^{-1}\mathrm{Mpc}$ and for the pair orientation $\mu$ we use $100$ linear bins in the range $\mu=[0,1]$. In Fig.\ \ref{fig:zeff} we show the measured $\zeff(r,\mu)$ in the case of the NGC. While on smaller scales ($r<100\, h^{-1}\mathrm{Mpc}$) $\zeff$ shows no $\mu$-dependence, on large separations there is a clear dependence on $\mu$. The higher/lower $\zeff$ at lower/higher values of $\mu$ follows directly from the survey geometry --- at large separations there are more transverse pairs compared to the radial pairs.

\subsection{Clustering evolution}
\begin{figure}[!ht]
\includegraphics[width=0.48\textwidth]{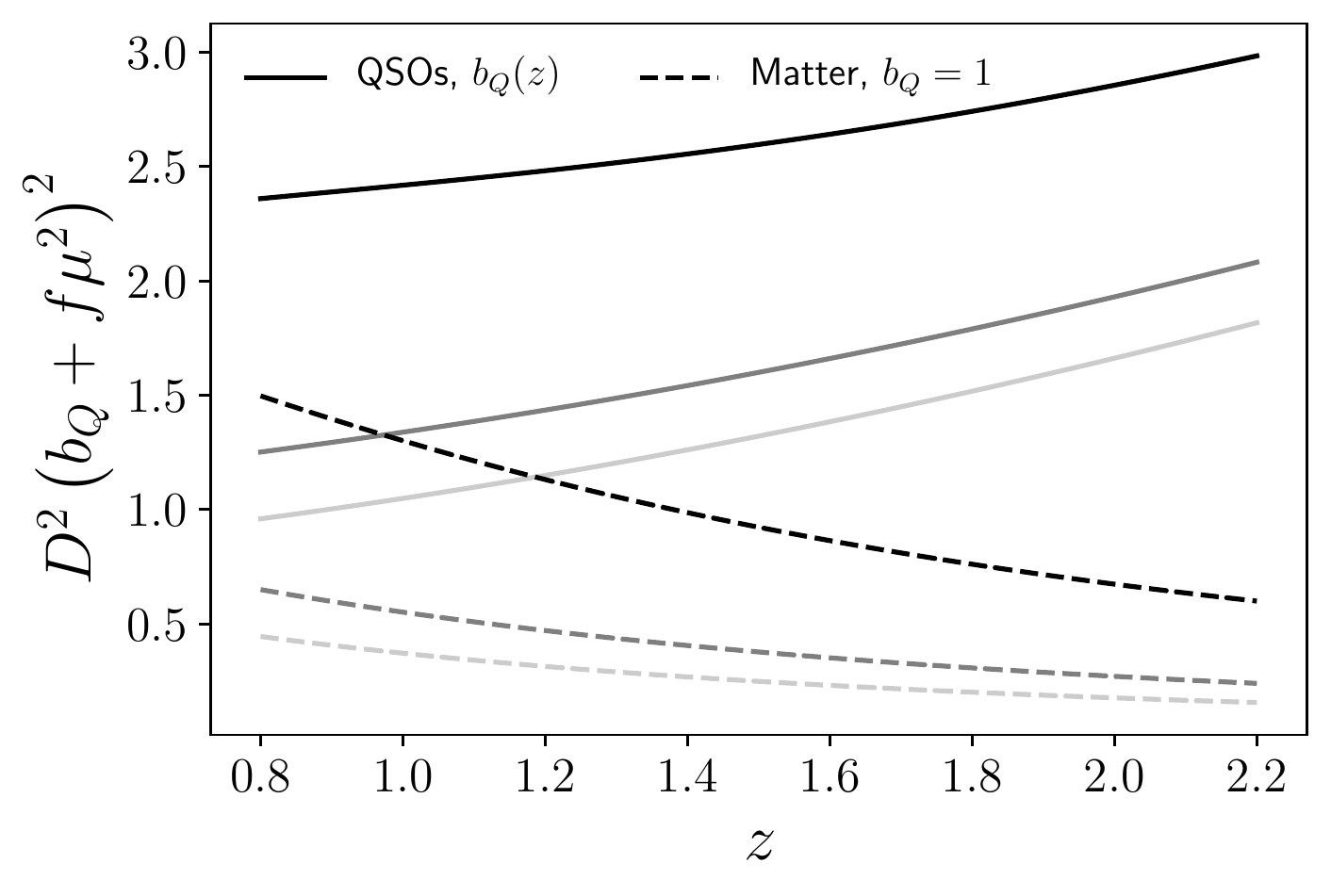}\\
\caption{The overall clustering strength evolution in linear RSD theory in the case of eBOSS QSOs (solid lines) and matter field (dashed lines) across the eBOSS QSOs redshift range. Different line shadings, from lightest to darkest, correspond to values of $\mu=\left[0,0.5,1\right]$, respectively.}
\label{fig:zevolution}
\end{figure}

The eBOSS QSOs are known to have a strong linear bias evolution. This bias evolution has been measured using the clustering multipoles from previous eBOSS data releases \cite{bQSO_Laurent,bQSO_Hector}. For the purposes of this paper, we adopt the parametrization and best-fit values from Ref.\ \cite{bQSO_Laurent}. Apart from the bias dependence, the overall clustering amplitude also contains the dependence on the growth factor $D(z)$ and the growth rate $f(z)$. Together, all of these three factors evolve differently across the redshift bin. In Fig.\ \ref{fig:zevolution} we show the redshift evolution of the clustering amplitude prefactor that contains all the redshift-dependent quantities across the QSOs $z$-range. In addition to QSOs, we also consider matter clustering only and show the evolution at several $\mu$-values. An isotropic effect of such redshift evolution of clustering is well known and has been studied in the literature \cite{Matarrese,Percival04,White_Martini_Cohn_QSO,2015MNRAS.447..234W,Vlah}. However, such evolution, coupled with the anisotropic $\zeff$, can have an effect on the interpretation of anisotropic clustering. In particular, while it may be negligible for the monopole, it can have larger effect on the higher clustering multipoles.

\section{Impact on clustering analysis}\label{results}
In this section we proceed to compare the linear theory clustering prediction obtained either keeping the $\zeff$ fixed or accounting for the $\zeff$ anisotropy.

In \S\ref{Effectivez} we showed that $\zeff$ additionally depends on $(r,\mu)$. This results in an additional $(r,\mu)$ dependency of the galaxy power spectrum prediction through the dependence of $b$, $f$ and $D$ on $(r,\mu)$. We first focus our analysis on the two-point correlation function $\xi(r,\mu)$ and we then translate the effect to the power spectrum multipoles.
\begin{figure}[!ht]
\includegraphics[width=0.48\textwidth]{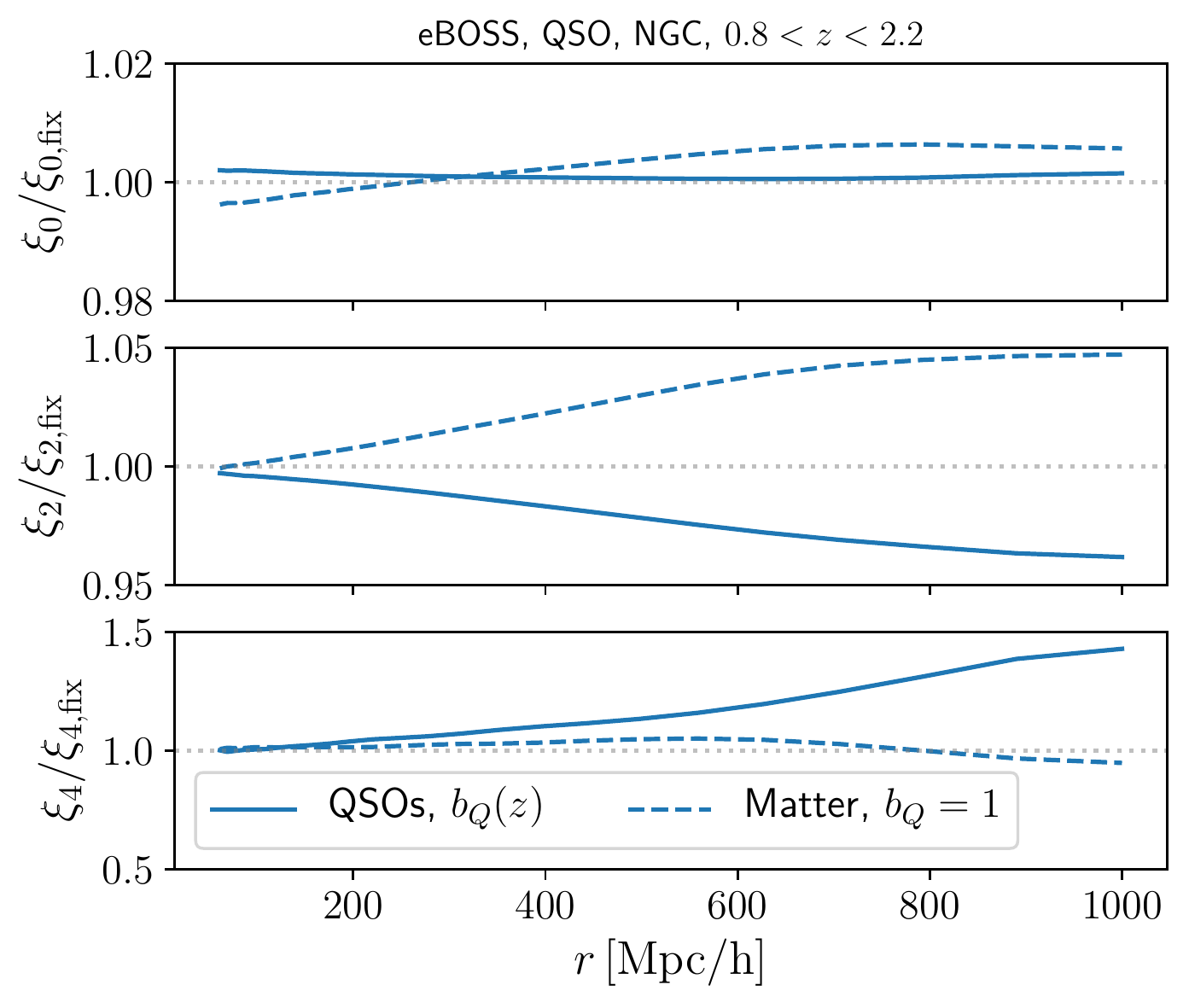}\\
\caption{The effect of including anisotropic effective redshift on clustering multipoles assuming linear theory compared to the case of fixed (isotropic) effective redshift. We show both the case of eBOSS QSO (solid line), i.e.\ biased tracers, and matter field (dashed line) across the same redshift range.}
\label{fig:xiell}
\end{figure}

We show the relative effect of including $\zeff$ on the correlation function multipoles in Fig.\ \ref{fig:xiell}. We compute the correlation function multipoles using Eq.\ \eqref{eq:xiell} by either using the measured $\zeff(r,\mu)$ in Eq. \eqref{eq:Pkmuzeff} or using a single effective redshift. In the later case we use $\zeff=1.47$ \cite{Hou}. While the effect on the monopole is sub-percent, the effect is evident for quadrupole and hexadecapole, reaching $\sim4\%$ and $\sim 40\%$, respectively, on the largest scales. Additionally, we repeat the same calculation in the absence of evolving galaxy bias. For this we consider the case of matter field ($b_Q=1$) having the same survey geometry as the QSOs. In Fig.\ \ref{fig:xiell} we show that additional large scale anisotropy is present even in the case of matter and comparable in amplitude to the considered case of biased tracers.

\begin{figure}[!ht]
\includegraphics[width=0.48\textwidth]{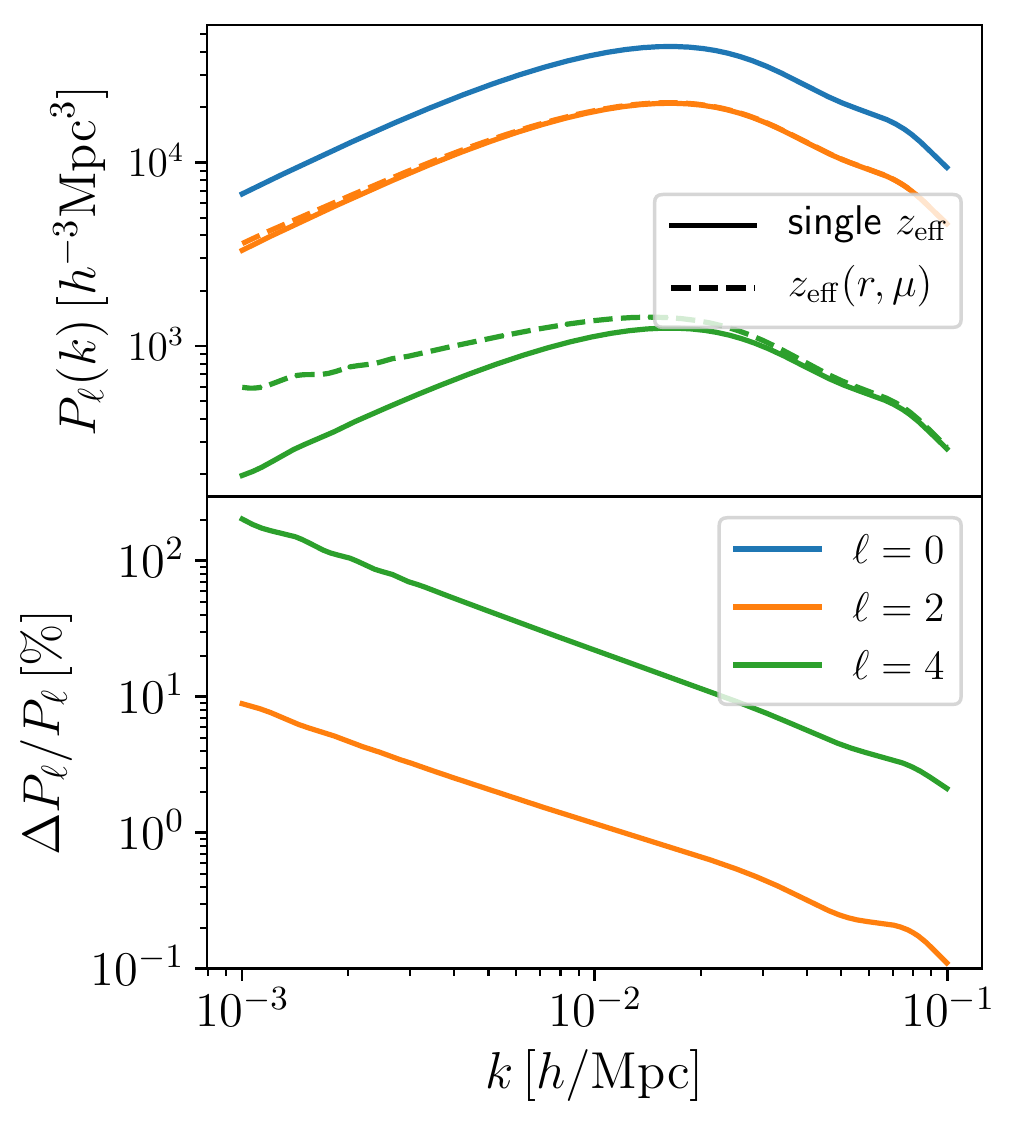}\\
\caption{\textit{Top panel:} The power spectrum multipoles computed assuming either a single $\zeff$ (solid lines) or an anisotropic $\zeff$ (dashed lines) in linear theory in the case of eBOSS NGC QSO. Different multipoles are shown in different colors in both panels (see legend in the bottom panel). \textit{Bottom panel:} The relative effect on the power spectrum multipoles as a function of scale. For simplicity, we do not include the window function effects on the multipoles.}
\label{fig:P_ell}
\end{figure}

Based on this analysis, we estimate the effect on the power spectrum multipoles. As the effect on the monopole is sub-percent on all scales, we only consider the effect on higher order multipoles. To do this we approximate the effect on the correlation function multipoles as follows:
\be
\xi_\ell^{\zeff} \approx \xi_\ell(r) \left[ 1 + \Delta_\ell(r/1000\, h^{-1}\mathrm{Mpc})\right],
\ee
where $\Delta_\ell\,[\%]=(5, 50)$ for $\ell=(2,4)$, respectively. We then transform $\xi_\ell^{\zeff}$ to $P_\ell^{\zeff}$ using Eq. \ref{eq:xiell} and show the results in Fig.~\ref{fig:P_ell}. As expected, the effect is more pronounced on larger scales (smaller $k$'s). The quadrupole and hexadecapole are affected at the level of $10\%$ and $100\%$, respectively, at scales of $k\sim 10^{-3}\,[h/\mathrm{Mpc}]$.
\begin{figure}[!ht]
\subfloat{\includegraphics[width=0.48\textwidth]{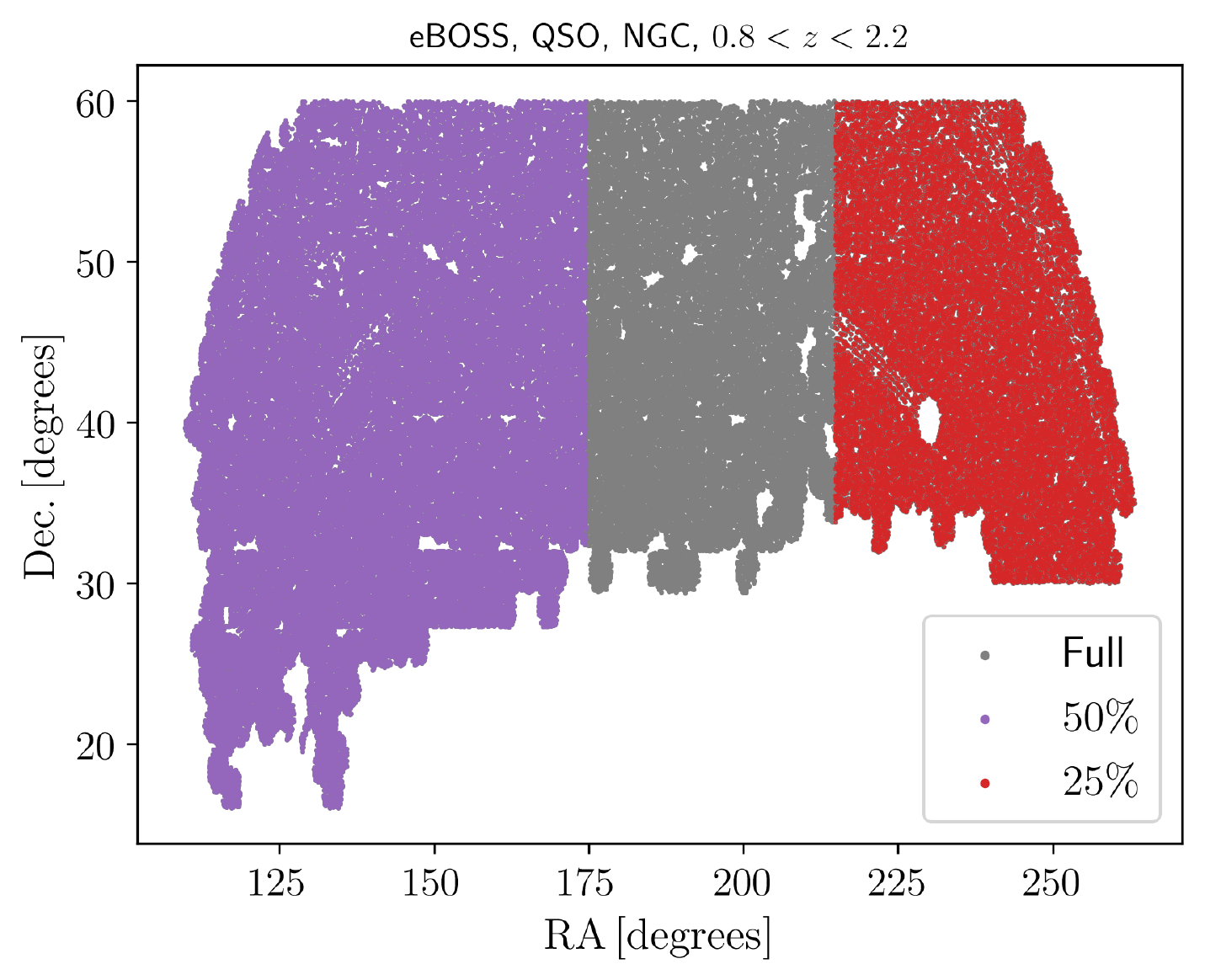}}\\
\subfloat{\includegraphics[width=0.48\textwidth]{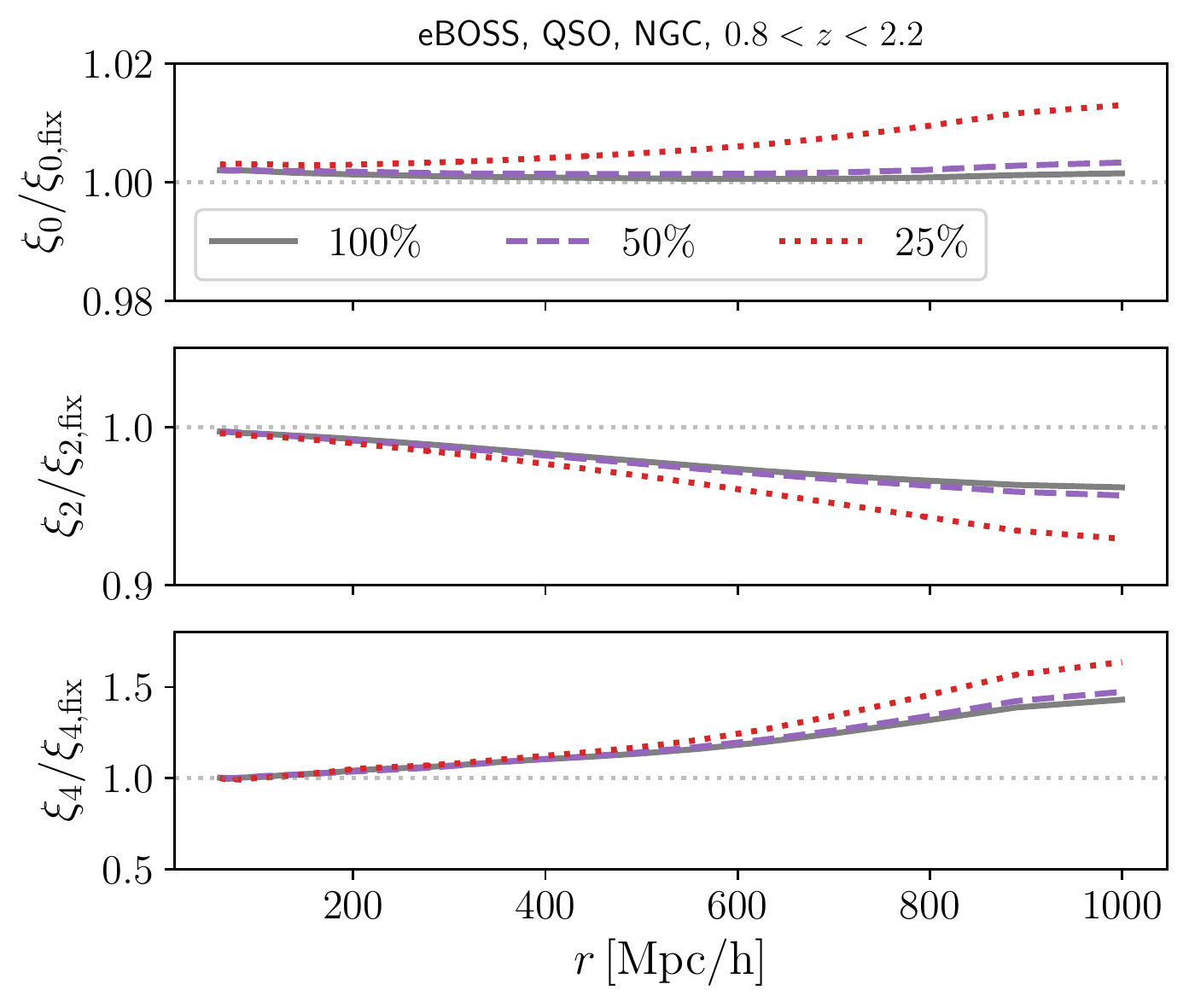}}
\caption{\textit{Top panel:} The full eBOSS NGC QSOs footprint (gray) along with the half (purple) and quarter (red) subsamples overplotted. For the purposes of this figure we display every 100-th datapoint from the original samples. \textit{Bottom panel:} Similar to Fig.\ \ref{fig:xiell}, we show the effect of the anisotropic $\zeff$ in the subsamples from the top panel (with matching colors) using the the eBOSS QSO.}
\label{fig:xiell_sub}
\end{figure}

\subsection{Dependence on the survey area}
To demonstrate the geometrical nature of the $\zeff$ effect, we consider subsamples of the QSO catalogue, split into contiguous regions on the sky and estimate the effect on the clustering multipoles as a function of the sky area. We consider subsamples of the total eBOSS NGC QSO footprint based on the Right Ascension (RA). We select a half and a quarter of the full eBOSS NGC QSO sky area and we show these subsamples in Fig.\ \ref{fig:xiell_sub} (top panel). As these subsamples are chosen solely on angular positions, they have matching redshift distributions and clustering amplitude evolution within the redshift range. The only difference is in the geometry of the sample. For each of these subsamples we repeat the analysis of \S\ref{results} and compute the impact on the correlation function multipoles. We show the results in Fig.~\ref{fig:xiell_sub} (bottom panel). We find that the impact of anisotropic $\zeff$ has a stronger impact on the multipoles when smaller survey areas are considered. 

\subsection{Mitigating this effect}
We envision several ways to mitigate the effects that we have studied in this paper. One approach would be to weight pairs of galaxies by the reciprocal of their expected clustering strength before measuring the clustering. For each pair, we would model the ratio of the expected clustering strength for typical galaxies, divided by the expected clustering strength for those particular galaxies, and then weight each pair by this. This approach is similar to that adopted in \cite{Percival04}, in which each galaxy is weighted by $1/(D(z)\times \langle b\rangle)$, where $\langle b\rangle$ is the expected relative galaxy bias to the one of the full sample. This weighting was adopted to remove the isotropic component of the effect we are discussing here, i.e.\ they considered that pairs of different separation in the monopole tend to pick up different types of galaxy at different redshifts, leading to offsets in the monopole. Their analysis did not consider anisotropy in the window and the effect on quadrupole and hexadecapole. We have now shown that the effect on the anisotropic clustering from this coupling of the window and the clustering strength is actually stronger than the isotropic effect. Consequently, weighting each galaxy by the reciprocal of the expected clustering strength for that galaxy would not work: we need to apply a pairwise weight that depends on the RSD signal to ensure that the effect is removed. In terms of redshift, this approach would bring all galaxy pairs to a common effective redshift which is both scale- and orientation-independent. This would then allow for the standard clustering analysis to be applied, although we would have to worry that the correction is model dependent, and so if the best-fit model if far from this we may need to iterate with another set of weights. 

Another approach would be similar to the one we have taken in this paper, i.e.\ to include the clustering evolution and the anisotropic $\zeff$ when computing the theoretical model. In a survey analysis, we would need to allow for the evolution in expected clustering strength with redshift in the model, before convolving with the window function.
\begin{figure}[!ht]
\includegraphics[width=0.48\textwidth]{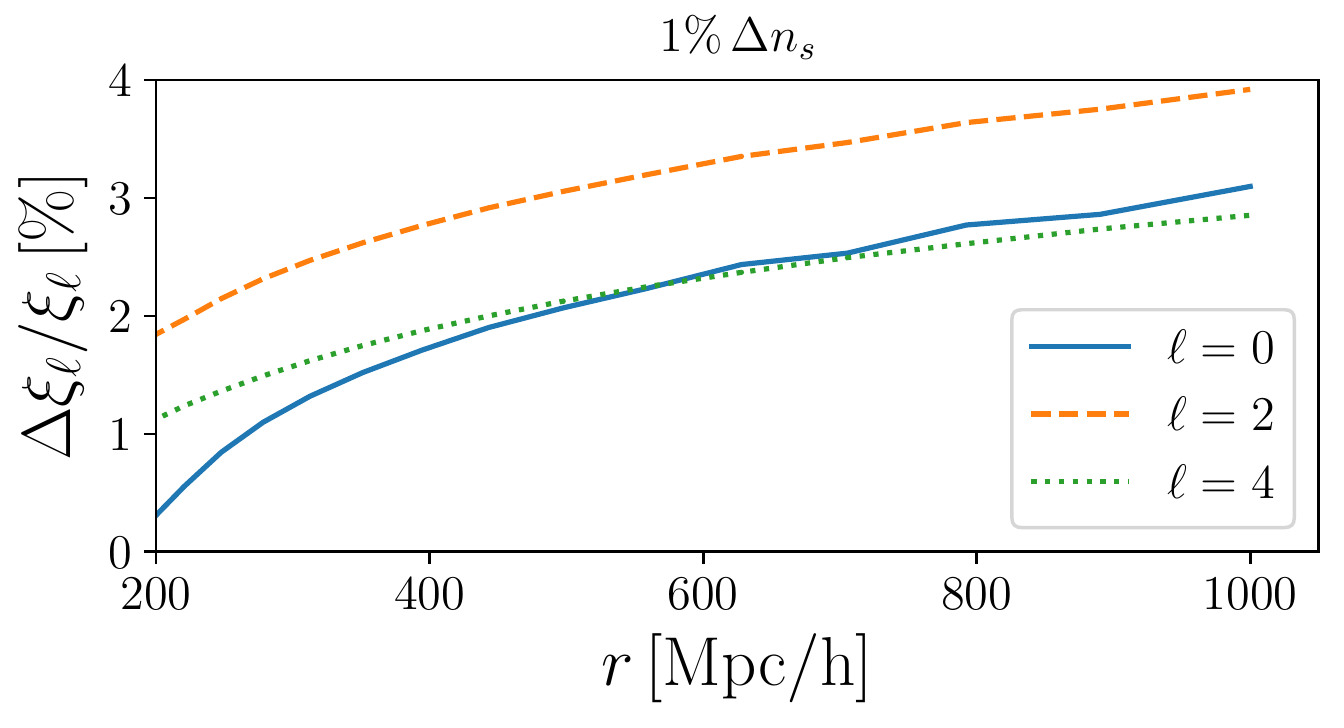}
\caption{The relative effect of changing $n_s$ by $1\%$ on the correlation function multipoles $\ell=0,2,4$ (see legend) on large-scales assuming linear theory.}
\label{fig:ns}
\end{figure}

\section{Discussion and Summary}\label{discussion}
In this paper we have considered the combination of two effects, which has previously not been considered in the clustering analysis of galaxy redshift surveys which span wide redshift ranges. One effect is due to the $\mu$-distribution of galaxy pair orientations, which are not isotropic at large separations due to the survey geometry. This makes the effective redshift defined at the pair level to be anisotropic, i.e.\ it depends on both the pair separation and orientation. While in the standard clustering analyses one compares the clustering measurements with the model computed at a single effective redshift, we have measured the effective redshift as a function of both separation and orientation and showed variations with pair orientation at large separations. In itself this would not be a problem if the amplitude of the clustering signal was the same throughout the survey. However if not, the anisotropic effective redshift couples with the clustering evolution and can result in an additional scale-dependent anisotropy in the measured clustering. The combined effect causes a large-scale tilt in the clustering multipoles. This points to a need to either account for the non-uniform $\mu$ distribution and the redshift evolution when computing the theoretical model, or correct for the effect at the level of clustering measurements.

The effect of anisotropic $\zeff$ on large-scale multipoles is strongly dependent on the survey geometry and clustering evolution of a particular sample. This makes it hard to make any general conclusions on the impact on cosmological parameters. However, being such a large-scale effect one can imagine parameters that could be most impacted. As an example, the tilt of the primordial power spectrum --- $n_s$, has a similar effect on the clustering multipoles. In Fig.\ \ref{fig:ns} we show the effect of a $1\%$ change in $n_s$ on the multipoles in linear theory. While varying $n_s$ has similar effect on quadrupole as the anisotropic $\zeff$, it affects all multipoles at a comparable level, whereas anisotropic $\zeff$ has very different effect on different multipoles. Thus, when considering all clustering multipoles the effects of anisotropic $\zeff$ and $n_s$ could be distinguished between each other. However, different survey geometries and/or samples with different clustering evolution could, in principle, result in comparable effect on all clustering multipoles due to anisotropic $\zeff$ and thus potentially be degenerate with the effects of $n_s$.

Another important effect is the presence of PNG which leave a distinct scale-dependent bias on large scales. While PNG leave the strongest effect on the monopole, as it is most sensitive to the galaxy bias, the effect of $\zeff$ is stronger for higher order multipoles. Since the current PNG constraints from galaxy clustering are coming solely from the power spectrum monopole \cite{Castorina_fnl}, we expect $\zeff$ to have a negligible effect. However, future analyses could start exploiting the PNG information from the higher multipoles as well, where it may become important to account for the effects of anisotropic $\zeff$ and clustering evolution.

Other approaches have been proposed for analysing surveys spanning wide redshift ranges and they include: splitting the sample into multiple redshift-bins, binning at the level of pair centers \cite{Nock_pair_centers} or using optimal redshift weighting \cite{Zhu_zweights,Ruggeri_zweights,Castorina_fnl}. However, these approaches will not help with the effects we have presented here. Splitting into multiple redshift-bins is generally sub-optimal, as it loses the information from pairs crossing the bin boundaries, and even though it is possible to obtain the large scales by cross-correlating different bins, one still needs to correct for the anisotropic pair distribution. While the second approach still contains large separation pairs, the anisotropic distribution and the window functions cares about the full distribution of pairs, not just their centres. The weighting introduced by the third approach is not designed to correct the anisotropic pair distribution, and we therefore expect this to be important after weighting.

While we have mainly focused on eBOSS QSOs as a specific dataset to demonstrate the combination and the size of these effects, this combination of effects is expected be present at some level for other galaxy samples as well. With the required precision of the upcoming galaxy surveys such as DESI and Euclid in mind, effects like this are expected to become increasing more important for future clustering analysis. 

\begin{acknowledgments}
AO thanks Emanuele Castorina, Francesca Lepori and Marko Simonovi\'{c} for useful discussions. We acknowledge support provided by Compute Ontario (www.computeontario.ca) and Compute Canada (www.computecanada.ca). We also acknowledge the use of \texttt{nbodykit} \cite{nbodykit}, \texttt{IPython} \cite{IPython}, \texttt{Matplotlib} \cite{Matplotlib}, \texttt{NumPy} \cite{Numpy2020} and \texttt{SciPy} \cite{SciPy}.

Funding for the Sloan Digital Sky 
Survey IV has been provided by the 
Alfred P. Sloan Foundation, the U.S. 
Department of Energy Office of 
Science, and the Participating 
Institutions. 

SDSS-IV acknowledges support and 
resources from the Center for High 
Performance Computing  at the 
University of Utah. The SDSS 
website is www.sdss.org.

SDSS-IV is managed by the 
Astrophysical Research Consortium 
for the Participating Institutions 
of the SDSS Collaboration including 
the Brazilian Participation Group, 
the Carnegie Institution for Science, 
Carnegie Mellon University, Center for 
Astrophysics | Harvard \& 
Smithsonian, the Chilean Participation 
Group, the French Participation Group, 
Instituto de Astrof\'isica de 
Canarias, The Johns Hopkins 
University, Kavli Institute for the 
Physics and Mathematics of the 
Universe (IPMU) / University of 
Tokyo, the Korean Participation Group, 
Lawrence Berkeley National Laboratory, 
Leibniz Institut f\"ur Astrophysik 
Potsdam (AIP),  Max-Planck-Institut 
f\"ur Astronomie (MPIA Heidelberg), 
Max-Planck-Institut f\"ur 
Astrophysik (MPA Garching), 
Max-Planck-Institut f\"ur 
Extraterrestrische Physik (MPE), 
National Astronomical Observatories of 
China, New Mexico State University, 
New York University, University of 
Notre Dame, Observat\'ario 
Nacional / MCTI, The Ohio State 
University, Pennsylvania State 
University, Shanghai 
Astronomical Observatory, United 
Kingdom Participation Group, 
Universidad Nacional Aut\'onoma 
de M\'exico, University of Arizona, 
University of Colorado Boulder, 
University of Oxford, University of 
Portsmouth, University of Utah, 
University of Virginia, University 
of Washington, University of 
Wisconsin, Vanderbilt University, 
and Yale University.
\end{acknowledgments}

\bibliography{References}
\bibliographystyle{JHEP}
\end{document}